\documentclass[aps,prl,a4paper,superscriptaddress,amsmath,amssymb,twocolumn,showpacs]{revtex4}
\usepackage{graphicx,epsfig}
\usepackage[usenames]{color}
\usepackage{amsfonts}
\usepackage{amsmath}
\usepackage{amssymb}

\providecommand{\U}[1]{\protect\rule{.1in}{.1in}}

\begin{document}

\title{Thermodynamics of the $\alpha$-$\gamma$ transition in cerium studied by an LDA + Gutzwiller method}

\author{Ming-Feng Tian}
\affiliation{Data Center for High Energy Density Physics, Institute of Applied
Physics and Computational Mathematics, Beijing 100088, China}
\affiliation{Beijing National Laboratory for Condensed Matter Physics and
Institute of Physics, Chinese Academy of Sciences, Beijing 100190, China}
\affiliation{Software Center for High Performance Numerical Simulation, China
Academy of Engineering Physics, Beijing 100088, China}

\author{Hai-Feng Song}
\affiliation{Data Center for High Energy Density Physics, Institute of Applied
Physics and Computational Mathematics, Beijing 100088, China}
\affiliation{Software Center for High Performance Numerical Simulation, China
Academy of Engineering Physics, Beijing 100088, China}

\author{Hai-Feng Liu}
\affiliation{Data Center for High Energy Density Physics, Institute of Applied
Physics and Computational Mathematics, Beijing 100088, China}
\affiliation{Software Center for High Performance Numerical Simulation, China
Academy of Engineering Physics, Beijing 100088, China}

\author{Cong Wang}
\affiliation{Data Center for High Energy Density Physics, Institute of Applied
Physics and Computational Mathematics, Beijing 100088, China}

\author{Zhong Fang}
\affiliation{Beijing National Laboratory for Condensed Matter Physics and
Institute of Physics, Chinese Academy of Sciences, Beijing 100190, China}

\author{Xi Dai}
\affiliation{Beijing National Laboratory for Condensed Matter Physics and
Institute of Physics, Chinese Academy of Sciences, Beijing 100190, China}

\date{\today}


\begin{abstract}
The $\alpha$-$\gamma$ transition in cerium has been studied in both zero and finite temperature by
Gutzwiller density functional theory. We find that the first
order transition between $\alpha$ and $\gamma$ phases persists to the zero temperature with negative pressure.
By further including the entropy contributed by both electronic quasi-particles and lattice vibration, we obtain
the total free energy at given volume and temperature, from which we obtain the $\alpha$-$\gamma$ transition
from the first principle calculation. We also computed the phase diagram and pressure versus volume
isotherms of cerium at finite temperature and pressure, finding excellent agreement with
the experiments. Our calculation indicate that both the electronic entropy and lattice vibration entropy
 plays important role in the $\alpha$-$\gamma$ transition.
\end{abstract}

\maketitle

\bigskip

Once a material contains \textit{f}-electrons sitting on the blink of itineracy to localization  transition,
its lattice and electronic degree of freedom will be strongly coupled to each other
leading to fascinating structural phase diagrams. The most famous examples are the
structural phase transitions with large volume changes in metal plutonium and
cerium.
Comparing with the situation of plutonium, the cerium metal is much simpler because it has only
one 4\textit{f} electron, which makes it an "prototype material" to study the structural
phase transitions in \textit{f} electron materials.

Among the phase transitions in cerium metal, enormous research interests have been attracted to
the $\alpha$ to $\gamma$ transition, which is an iso-structural transition with volume collapse as large
as $14\%$ at room temperature and pressure around 0.8 GPa \cite{Koskenmaki_1978, Lipp_2008, Decremps_2011}).
The transition pressure rises with increasing temperature, and seems to end at a solid-solid
critical point (CP) around P$_c$ = 1.5 GPa and T$_c$ slightly under 500 K
\cite{Lipp_2008, Decremps_2011}. The nature of this transition has still under
debate and become one of the classical problems in condensed matter physics.

The key issues under debating can be summarized into the following three aspects.
i) Although it is  commonly believed that the electronic states of 4\textit{f} orbitals undergo
a dramatic change from $\alpha$ to $\gamma$ phases, whether this change can be better
described by Mott transition model \cite{Johansson_1974,Johansson_1975,Johansson_1995}
or Kondo volume collapse (KVC) model \cite{Allen_1982,Lavagna_1982,Allen_1992} is
still under debating \cite{Lipp_2008,Johansson_2009,Jeong_2004,Rueff_2006}. The main difference
between the above two models is the role of the $spd$ bands. In Mott transition model, these $spd$
bands are nearly "spectators" of the transition and the 4f bands become completely localized
in the $\gamma$ phase. While in the view point of KVC, there are no qualitative difference
between $\alpha$ and $\gamma$ phases. The only difference is the scale of
Kondo temperature. ii) Since the $\alpha$ to $\gamma$ transition also happens at
finite temperature with ambient pressure, it is quite clear that entropy difference is one of the
important driving force for the transition. The question is whether the transition is purely driven by entropy?
In another words, whether or not such a transition can also be induced at the zero temperature
(for example by negative pressure), where entropy plays no role. iii) Given the fact
that entropy is important to the transition, what is its origin? Is it mainly contributed
by electronic entropy or lattice entropy?

Besides the experimental studies
\cite{Beecroft_1960,Lipp_2008,Wuilloud_1983,Wieliczka_1984,
Jeong_2004,Rueff_2006,VanderEb_2001}, the first principle calculation is another powerful tool
with parameter free to study the nature of the structural phase transitions.
In the density-functional theory (DFT) calculations with local density approximation (LDA) or
general gradient approximation (GGA), the correlation effects
among the \textit{f} electrons have not been fully considered in a satisfactory way,
leading to overestimation of the kinetic energy for the \textit{f} electrons. Therefore as one of the consequences,
only the $\alpha$ phase of cerium can be obtained and there is no signal for the appearance
of $\gamma$ phase at all even after including both the electronic and
vibrational entropy
\cite{Jarlborg_1997,YiWang_2000,Luders_2005,YiWang_2008}.
While on the other hand, the self-interaction corrected local spin density approximation (LSDA)
\cite{Luders_2005,Szotek_1994,Svane_1994} and LSDA + U calculations can obtain the $\gamma$ phase
by assuming the 4\textit{f} electrons to be
 either completely localized or magnetic, but still it is difficult
for these methods to describe both phases under a unified physical picture.
Besides, a calculation base on hybrid density functionals
obtain two distinct solutions at zero temperature that can be associated with
the $\alpha$ and $\gamma$ phases of Ce \cite{Casadei_2012}, but the results have not yet
extended to the finite temperature case. Nevertheless, a recent work from
density functional theory proposed that thermal disorder contributes via entropy to the stabilization of the
$\gamma$ phase at high temperature \cite{Jarlborg_2014}, the $\alpha$-$\gamma$ transition is
calculated to occur around 600 K.

Within the LDA + DMFT method, a combination of  LDA with dynamical mean field theory
(DMFT), early numerical studies have been carried out to study
the phase transitions of Ce at finite temperature
\cite{Zolfl_2001,Held_2001,McMahan_2003,Haule_2005,McMahan_2005,Amadon_2006,Chakrabarti_2014}. While since the
quantum Monte Carlo methods have been adopted as the impurity solver of DMFT, it is difficult
for LDA + DMFT to study the transition in low temperature and the full thermodynamic
features of the transition in the full temperature range have yet to be obtained.

In this paper we show that LDA + Gutzwiller method, which
incorporate LDA with Gutzwiller variational approach, can be well applied to study the
ground properties of the cerium metal and with the generalization to the finite
temperature, it can be further applied to study the thermodynamic properties of the $\alpha$
to $\gamma$ transition in low temperature. Here we will sketch the most
important aspects of the method, and leave the details to Refs. \cite{xydeng_2008, xydeng_2009, Ho_2008}.
The total Hamiltonian to describe the strongly correlated systems can be written as
\begin{equation} \label{eq:H}
H =H_{LDA}+H_{int}-H_{DC},
\end{equation},
where $H_{LDA}$ is the single particle Hamiltonian obtained
by LDA and $H_{int}$ is the local interaction term for the 4\textit{f}
electrons. $H_{DC}$ is the double counting
term representing the interaction energy already considered
at the LDA level. In the present paper, we compute
the double counting energy using the scheme described in
Ref. \cite{xydeng_2009}. In LDA + Gutzwiller we apply the following Gutzwiller trial wave
function,  $|G\rangle=P_{G} |0\rangle$, where $P_{G}$ is the Gutzwiller projectors containing
variational parameters to be optimized by the variational principle and the non-interacting
state $|0\rangle$ is the solution of the effective Hamiltonian for the quasi-particles
$H_{eff} \approx P_GH_{LDA}P_G$. The ground state properties of cerium metal have been
studied using LDA + Gutzwiller by us for the positive pressure case \cite{mftian_2011}
and recently by G. Kotliar's group for the negative pressure case
\cite{Lanata_2013}, where they find that for interaction strength
$U < 5.5 eV$ the first order transition between
$\alpha$ and $\gamma$ phases survives at zero temperature, they also reported
a new implementation of the Gutzwiller approximation
\cite{Lanata_2014a,Lanata_2014b,Lanata_2014c}. Recently, they implement an
LDA + SB method studied the $\alpha$-$\gamma$ phase transition and phase diagram
of cerium at finite temperature \cite{Lanata_2014d}, and obtain
results in good agreement with the experiments.

In the present paper, we apply the LDA + Gutzwiller method implemented in our
pseudo potential plane wave code BSTATE \cite{GTWang_2008, xydeng_2008, xydeng_2009,
mftian_2011} to obtain the ground state energy of cerium metal crystallized in face
centered cubic (fcc) structure.  The LDA part of the calculations were performed with
the full consideration of the  relativistic effect and a
16$\times$16$\times$16 k mesh for higher energy converged precision.

\begin{figure}[!ht]
\includegraphics[width=8cm]{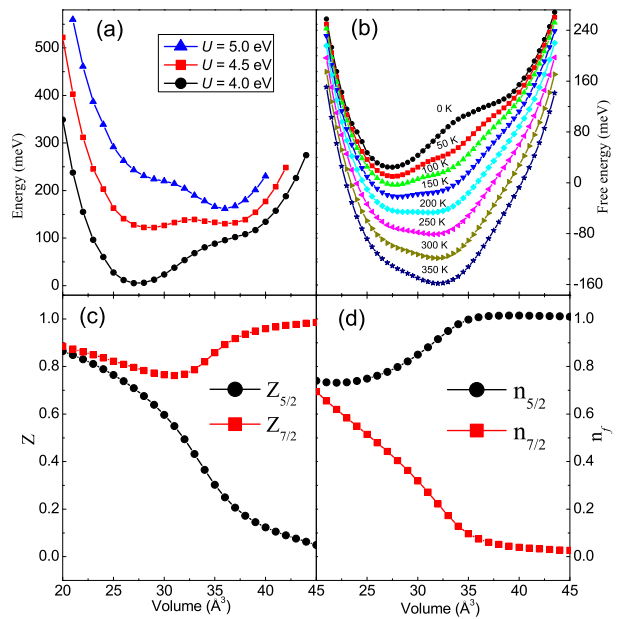}
\caption{(Color online) Calculate cold energy for different \textit{U} (upper
left panel), and free energy for different temperature (upper right panel) as
a function of atomic volume. Quasiparticle renormalization weights of the 7/2
and 5/2 \textit{f}-electrons (lower left panel), 7/2 and 5/2 orbital
populations (lower right panel).\label{fig1}}
\end{figure}

Our main results in zero temperature are shown in Fig. \ref{fig1}(a). The negative
curvature region in the total energy versus volume curve, which signals the first order
transition with pressure, is present for all interaction strength, which is slightly
different with the results obtained by G. Kotliar's group\cite{Lanata_2013}.
With interaction strength
$U = 4.0 eV$, both the experimental volume (28.0 - 29.0 \AA$^3$ ) and
bulk modulus (20.0 - 35.0 GPa) \cite{Ellinger_1974, Zachariasen_1977, J.S.Olsen_1985}
at ambient pressure can be nicely reproduced and we will adopt this value for the calculations through
out the paper.  The quasi-particle weight and the
average occupation of both the $j=5/2$ and $7/2$ bands are plotted in
Fig. \ref{fig1}(c) and (d). From Fig. \ref{fig1}(d) one can find that
the occupation number  of $j=5/2$ bands increase dramatically from 0.7 to almost
one in the volume regime from 29.0 \AA$^3$ (the equilibrium volume of the $\alpha$
phase)to 35.0\AA$^3$ (the equilibrium volume of the $\gamma$ phase)and at the same
time the quasi-particle weight drops abruptly (Fig. \ref{fig1}(c)), indicating that the \textit{f}-electrons
is quite itinerant in $\alpha$ phase while becomes quite localized in $\gamma$ phase,
which can be better described by the Kondo lattice model.

In order to directly compare the first principle results with the experimental data, which is always performed at finite
temperature, the calculations of the free energy at finite temperature is strongly desired. In the present paper, we generalize
the LDA + Gutzwiller method to calculate the free energy by including both the
electronic and lattice vibrational entropy. The total
Helmholtz free energy can be always written as
\begin{equation} \label{eq:Free_enegy}
F(V,T)=F_{vib}(V,T)+F_{el}(V,T)
\end{equation}
where $F_{el}(V,T)$ and $F_{vib}(V,T)$ are the electronic and lattice
vibrational part of the free energy respectively.
In the present study, we assume that at least in the low temperature both the $\alpha$ and
$\gamma$ phases are in the fermi liquid region, where the electronic entropy can be calculated
by counting the thermally excited quasi-particles. Near the critical temperature, which is around
500K, the $\gamma$ phase will be no longer in the fermi liquid phase any more, leading to
possible underestimation of the electronic entropy in $\gamma$ phase, which will be discussed
in more detail below.

Therefore in the present study, the electronic free energy will be estimated as
$$
F_{el}(V,T)=\int{n_{qp}(\epsilon,V)f(\epsilon)\epsilon d\epsilon}+
$$
\begin{equation} \label{eq:F_el}
Tk_{B}\int{n_{qp}(\epsilon,V)[f\texttt{ln}f+(1-f)\texttt{ln}(1-f)]
d\epsilon}
\end{equation},
where $n_{qp}(\epsilon,V)$ is the quasi-particle density of states obtained by solving the Gutzwiller effective Hamiltonian
$H_{eff}$ and $f(\epsilon)$ denotes the Fermi-Dirac distribution function. The first and second terms in the above equation
denote the energy and entropy contributions to the electronic free energy respectively.

The lattice part of the free energy is estimated within the mean field approximation proposed in the references
\cite{Cohen_1996,YiWang_2000,Songhf_2007}, where the vibrating motion of the cerium atoms can be approximately treated as independent three dimensional oscillators moving under the harmonic potential formed by all the surrounding atoms.
The strength of such mean field potential can be approximately determined by the curvature of the
total energy versus volume curve as explained in detail in reference \cite{Cohen_1996,YiWang_2000,Songhf_2007}.
Considering the heavy mass of the cerium atoms, we can further treat the atomic motion classically and get the lattice free energy as
\begin{equation} \label{eq:F_ion}
F_{vib}(V,T)=-k_{B}T(\frac{3}{2}\texttt{ln}\frac{mk_{B}T}{2\pi\hbar^2}
+\texttt{ln}\upsilon_{f}(V,T))
\end{equation}
where
\begin{eqnarray} \label{eq:vf}
\upsilon_f(V,T)&=&4\pi \int{\texttt{exp}(-\frac{g(r,V)}{k_BT})r^2dr} \\
g(r,V)&=&\frac{1}{2}[E_{c}(R+r)+E_{c}(R-r)-2E_{c}(R)]
\end{eqnarray}
where \textit{r} represents the distance that the lattice ion deviates from
its equilibrium position, \textit{R} is the lattice constant, and $V=R^3/4$
in the case of fcc crystal.

The free energy curves with different temperature in Fig. \ref{fig1}(b)
reveals the competition between the $\alpha$ and $\gamma$ phases,
the free energy difference between $\gamma$ phase and $\alpha$ phase decreased
with increasing temperature, which becomes almost zero at a temperature
of 190 K. This results illustrate that the $\alpha$-$\gamma$
transition temperature is 190 K at zero pressure, which agrees well with the
experimental data. \cite{Koskenmaki_1978}.

\begin{figure}[!ht]
\includegraphics[width=0.45\textwidth]{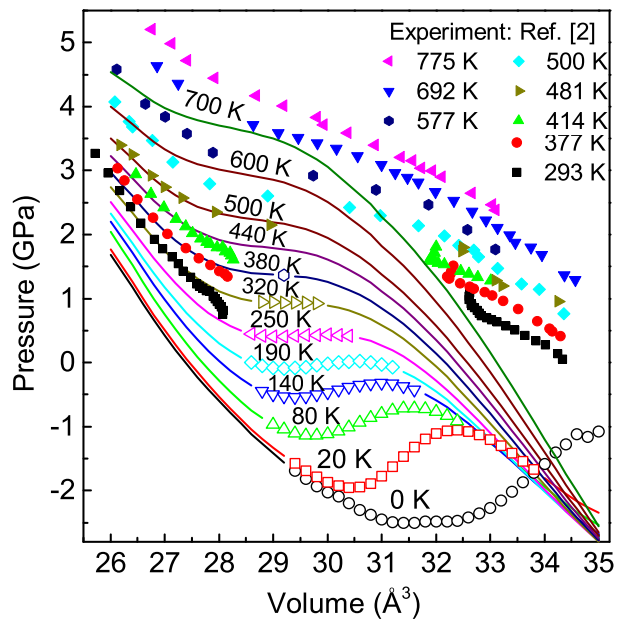}
\caption{(Color online) Pressure versus volume isotherms of fcc Ce. The solid line and open symbols are
results of LDA + G calculations, where the open symbols are the region of thermodynamic instability, i.e., the
$\alpha$-$\gamma$ transition. The solid symbols are previous experimental data \cite{Lipp_2008}. \label{fig2}}
\end{figure}

The pressure at given volume and temperature can be estimated as
$P(V,T)=-\partial{F(V,T)/\partial{V}}$, we calculate the pressure versus volume
isotherms of fcc Ce, as shown in
Fig. \ref{fig2}. At given pressure, the first order phase transition can be signaled
by the appearance of multiple solutions with different volumes, which has been plotted
in the figure by the open symbols indicating the region of thermodynamic instability.
Our isotherms of $\alpha$-Ce (volume little than thermodynamic instability region)
are excellent consistent with experimental data \cite{Lipp_2008}. Due to the lack of
magnetic entropy in LDA + G method, our results of $\gamma$-Ce are slightly
underestimate than the experiments, we will discuss this issue again in this paper later.

\begin{figure}[!ht]
\includegraphics[width=0.45\textwidth]{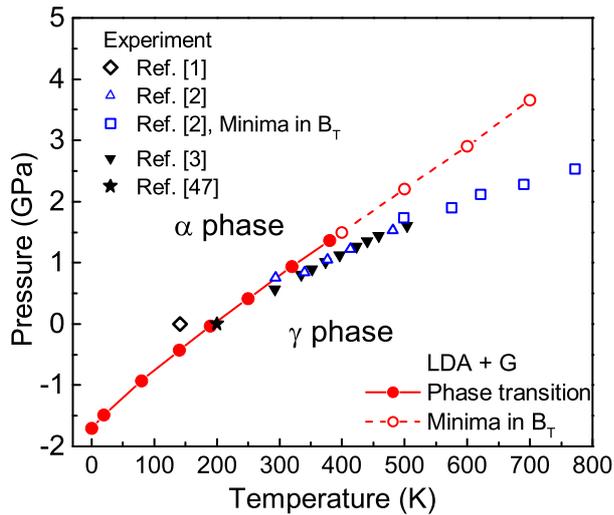}
\caption{(Color online) Pressure versus temperature phase diagram of fcc Ce.
Open and close ciecles denote present LDA + G data. The other symbles are
previous
experimental data ({\large $\diamond$} \cite{Koskenmaki_1978}, $\triangle$
\cite{Lipp_2008},
$\square$ \cite{Lipp_2008}, {\large $\blacktriangledown$}
\cite{Decremps_2011},
$\bigstar$ \cite{Gschneidner_1977}).  \label{fig3}}
\end{figure}

The calculated phase diagram has been plotted in Fig. \ref{fig3}
together with the comparison to the experimental data from several different papers.
The agreement between our LDA + Gutzwiller calculation and the experimental data is surprisingly
well.

\begin{figure}[!ht]
\includegraphics[width=0.45\textwidth]{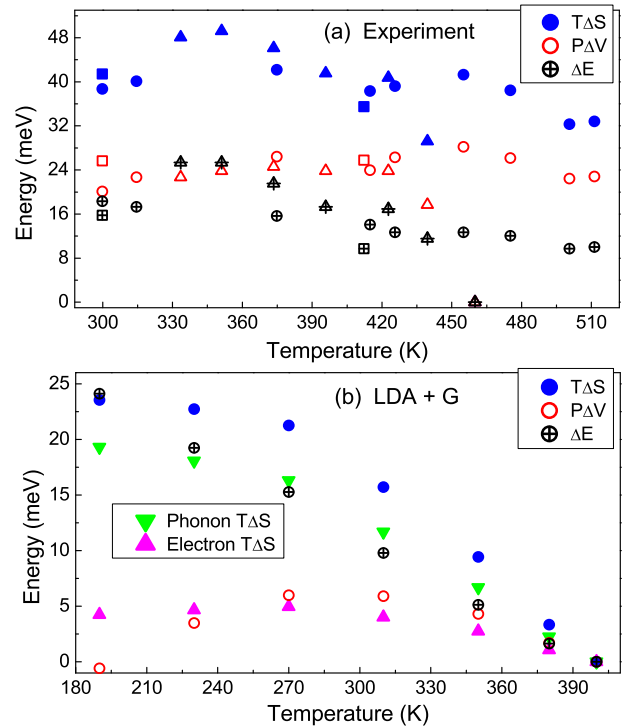}
\caption{(Color online) Energy related term across the
$\alpha$-$\gamma$ transition, upper panel are experimental data (${\bigcirc}$
\cite{Beecroft_1960}, $\square$ \cite{Schiwek_2002},
$\triangle$ \cite{Decremps_2011}), lower panel are present
data ($\bigtriangleup$ electronic entropy, $\bigtriangledown$ entropy of
phonon).
Entropy term T$\Delta$S (solid symbols), P$\Delta$V (open symbols), and
internal energy $\Delta$E (open symbols overlaps with plus).
\label{fig4}}
\end{figure}

Whether the $\alpha$ to $\gamma$ transition is mainly driven by entropy is another key question.
In this paper, we try to address it by comparing the change of three parts of
Gibbs free energy ($G(P,V,T)=F(V,T)+PV$) upon the transition,
T$\Delta$S, P$\Delta$V and internal energy $\Delta$E,
which are plotted in Fig. \ref{fig4}(b) together with the experimental results
in Fig. \ref{fig4}(a)
taken from reference (\cite{Beecroft_1960},\cite{Schiwek_2002}, \cite{Decremps_2011}). Our
results show that the biggest contribution to the transition comes from the
entropy change which is quite consistent with the experimental data
\cite{Beecroft_1960,Schiwek_2002,Decremps_2011}. We can further separate the entropy
contribution into electronic and lattice parts, which is also illustrated in
Fig. \ref{fig4}(b). The electronic
entropy change obtained by our LDA + Gutzwiller calculation is about 5.0 $meV$/atom, which is about 2-3 times
smaller than the lattice part. This is mainly due to the fact that in LDA + Gutzwiller only the quasi-particle
entropy has been included but not the magnetic entropy coming from the incoherent motion of the
\textit{f}-electrons. Considering the estimated Kondo temperature for $\gamma$ phase is about $70$ K
\cite{Rueff_2006}, the magnetic entropy, may also make sizable contribution to the $\gamma$ phase
and make the change of electronic entropy to be compatible to the lattice one upon the transition.

In conclusion, the thermodynamic features of the cerium $\alpha$ to $\gamma$ transition
has been obtained by applying the LDA + Gutzwiller method, from which we got the phase diagram
and isotherms of cerium, both in good agreement with the experiments. Our calculations also show that
the long puzzled transition is persists to the zero temperature with negative pressure,
and mainly driven by the entropy change, where both the electronic and the lattice part
play important roles.

This work was supported by the National Science Foundation of China (Grants
No. NSFC 11204015), by the 973 program of China (No. 2011CBA00108 and 2013CBP21700),
and by the Foundation for Development of Science and Technology of China Academy of
Engineering Physics (Grant No. 2011B0101011). We acknowledge the helpful discussion
with professor G. Kotliar and Dr. N. Lanat{\'a}.


\end{document}